\documentclass[aps,prb,floatfix]{revtex4}

\usepackage{graphicx}
\usepackage{amssymb,amsfonts,amsmath}
\usepackage{epsf}
\usepackage{subfigure}
\usepackage{epstopdf}
\usepackage{mathrsfs}

\newcommand{\be}{\begin{equation}}
\newcommand{\ee}{\end{equation}}
\newcommand{\bea}{\begin{eqnarray}}
\newcommand{\eea}{\end{eqnarray}}

\begin{document}

\title{Self-Organization at the Nanoscale Scale in \\ Far-From-Equilibrium Surface Reactions and Copolymerizations}

\author{Pierre Gaspard}

\affiliation{Center for Nonlinear Phenomena and Complex Systems,\\
Universit\'e Libre de Bruxelles, Campus Plaine, Code Postal 231,
B-1050 Brussels, Belgium}

\begin{abstract}
An overview is given of theoretical progress on self-organization at the nanoscale
in reactive systems of heterogeneous catalysis observed by field emission microscopy techniques and at the molecular scale in copolymerization processes.
The results are presented in the perspective of recent advances in nonequilibrium
thermodynamics and statistical mechanics, allowing us to understand how nanosystems
driven away from equilibrium can manifest directionality and dynamical order.
\end{abstract}

\noindent{\footnotesize A. S. Mikhailov and G. Ertl, Editors \\
Proceedings of the International Conference ``Engineering of Chemical Complexity"\\ 
Berlin Center for Studies of Complex Chemical Systems, 4-8 July 2011}

\vskip 1 cm

\maketitle

\section{Introduction}

In macroscopic systems, self-organization arises far from equilibrium beyond critical thresholds where the macrostate issued from thermodynamic equilibrium becomes unstable and new macrostates emerge through bifurcations. Such bifurcations may lead to oscillatory behavior and spatial or spatio-temporal patterns, called dissipative structures because they are maintained at the expense of free-energy sources.\cite{P67,NP77,IE95}  Recent developments have been concerned with the complexity of such nonequilibrium behavior, particularly, in small systems of nanometric size down to the molecular scale.\cite{NP89,NN07,FE11,G10} The molecular structure of matter largely contributes to the complexity of natural phenomena by the multiplicity and the variety of chemical species and their possible specific actions. Furthermore, the microscopic degrees of freedom manifest themselves at the nanoscale as molecular and thermal fluctuations, which requires a stochastic description for the thermodynamic and kinetic properties of small systems.  Remarkably, great advances have been recently achieved, leading to a fundamental understanding of the emergence of dynamical order in fluctuating nonequilibrium systems, as overviewed in Section \ref{Foundations}. 

These advances allow us to bridge the gap between the microscopic and macroscopic levels of description, especially, in reactions of heterogeneous catalysis studied by field electron and field ion microscopy techniques (FEM and FIM, respectively).\cite{vTGN92,GLRBE94,EBGKWB94,VK96,KV08}  In such reactions, dynamical patterns are observed on metallic tips with a curvature radius of tens of nanometers. Accordingly, the crystalline surface is multifaceted so that adsorption, desorption, reaction and transport processes have various speeds on different facets. Moreover, the activation barriers are significantly modified by the high electric field present under FEM or FIM conditions, as calculated by quantum electronic {\it ab initio} and density functional theories.\cite{VKGK06,MG06,MGMVK08}  These non-uniform and anisotropic properties of the catalytic surface participate to the generation of nanopatterns in the far-from-equilibrium regimes of bistability and oscillations observed, in particular, during water formation on rhodium, as presented in Section \ref{Catalysis}.\cite{VBK01,VBK04,MGVK09a,MGVK09b,MGVK10}

Under nonequilibrium conditions, the emergence of dynamical order is already in action at the molecular scale during copolymerization processes.  Copolymers are special because they constitute the smallest physico-chemical supports of information.  Little is known about the thermodynamics and kinetics of information processing in copolymerizations although such reactions play an essential role in many complex systems, e.g. in biology.  In this context, the recent advances in the thermodynamics of stochastic processes are shedding light on the generation of information-rich copolymers, as explained in Section \ref{Copolymers}.\cite{AG08,J08,AG09}

The purpose of this contribution is to present an overview of these recent advances
on self-organization at the nanoscale in the perspective of future theoretical and experimental work on these topics, as discussed in the next sections.

\section{Fundamental aspects of nonequilibrium nanosystems}
\label{Foundations}

\subsection{Structure and function of nanosystems}

There exist many different processes and systems at the nanoscale: heterogeneous catalysis on nanoparticles,\cite{GLRBE94,EBGKWB94,VK96,KV08} electrochemical reactions on nanoelectrodes,\cite{GMK10} synthetic molecular machines,\cite{SG11,LSBFLD11} single enzymes,\cite{MJYKQX05} linear and rotary molecular motors,\cite{FHTSY95,NYYK97} DNA and RNA polymerases responsible for replication and transcription,\cite{GB06,Eid09} or  ribosomes performing the translation of mRNAs into proteins.\cite{UEAKFTP10}  Every nanosystem has a specific structure and acquires its function when driven out of equilibrium by some free-energy source.  In this regard, the structure and function of a nanosystem can be characterized in terms of its equilibrium and nonequilibrium properties, as shown in Table~\ref{tbl1}.

\begin{table}[ht]
\caption{Comparison between the equilibrium and nonequilibrium properties of nanosystems.} 
\vskip 0.2 cm
{\begin{tabular}{@{}ll@{}} \toprule
Equilibrium & Nonequilibrium\\
\colrule
zero affinities & non zero affinities\\
zero mean fluxes & non zero mean fluxes \\
zero entropy production & positive entropy production\\
no free-energy supply needed \ \ \ \ \  & free-energy supply required\\
detailed balancing & directionality\\
3D spatial structure & 4D spatio-temporal dynamics\\
structure & function\\
\botrule
\end{tabular}}
\label{tbl1}
\end{table}

A nanosystem can be in thermodynamic equilibrium with its environment at given temperature and chemical potentials, as it is the case for a catalytic surface in contact with a gaseous mixture at chemical equilibrium or for an enzyme in a solution also at chemical equilibrium.  In these equilibrium systems, the ceaseless movements of thermal and molecular fluctuations do not need the supply of free energy to persist.  In particular, the catalytic sites are randomly visited by adsorbates or substrates but, on average, there is no flux of matter or energy between the pools of reactants and products.  The ratio of partial pressures or concentrations remains at its equilibrium value fixed by the mass action law of Guldberg and Waage.  Any movement in one direction is balanced by the reversed movement according to the principle of detailed balancing.  At equilibrium, there is no energy dissipation and no entropy production.  From a statistical-mechanical viewpoint, the molecular architecture of the nanosystem can be characterized in terms of the average relative positions of its atoms but their average velocities are vanishing.  In this respect, the nanosystem has only a 3D spatial structure at equilibrium.

In contrast, a nanosystem in an environment containing a mixture which is not in chemical equilibrium for the reactions the nanosystem can catalyze will sustain non-vanishing fluxes of matter or energy.  These average movements are driven by the free-energy sources of the environment.  Energy is dissipated and entropy produced.  For instance, a F$_1$-ATPase molecular motor rotates in a specific direction if it is  surrounded by a solution containing an excess of ATP with respect to the products of its hydrolysis.\cite{NYYK97,AG06PRE,GG10}  Therefore, nonequilibrium nanosystems acquire an average directionality, which can be controlled by the external nonequilibrium constraints, and they perform a 4D spatio-temporal dynamics, which is the expression of their function.  Accordingly, the function of a nanosystem holds in the specific 4D dynamics that its 3D structure can developed when it is driven away from equilibrium under specific conditions.

\subsection{Out-of-equilibrium directionality of fluctuating currents}

The kinetics of nanosystems close to or far from equilibrium has recently known tremendous progress with the advent of time-reversal symmetry relations also called fluctuation theorems.  These results find their origins in the study of large-deviation properties of chaotic dynamical systems sustaining transport processes of diffusion.\cite{GN90,G05}  Several versions of such relations have been obtained for systems under transient or stationary nonequilibrium conditions.\cite{ECM93,GC95,C99,LS99,EHM09,J11}  A particular version of the fluctuation theorem concerns the nonequilibrium work on single molecules subjected to the time-dependent forces of optical tweezers or atomic force microscopy.\cite{LSBFLD11,CRJSTB05}  For nonequilibrium systems in stationary states, a general  fluctuation theorem has been proved for all the currents flowing across open stochastic or quantum systems by using microreversibility.\cite{AG04,AG06JSM,AG07JSP,AG07JSM,AG08JSM,AG09JSM,AGMT09}

Nanosystems can be driven out of equilibrium by several independent thermodynamic forces, also called affinities.\cite{DD36}  For isothermal systems, they are defined as
\be
A_{\gamma} = \frac{\Delta G_{\gamma}}{k_{\rm B} T} \qquad (\gamma=1,2,...,c)
\label{affinity}
\ee
in terms of the Gibbs free-energy differences $\Delta G_{\gamma}=G_{\gamma}-G_{{\gamma},{\rm eq}}$ supplied by the nonequilibrium environment to power the mean motion.  They are external control parameters that depend on the concentrations or partial pressures of reactants and products.  A nanosystem between reservoirs at different temperatures and chemical potentials is characterized by thermal as well as chemical affinities.\cite{AGMT09}  

The affinities drive the currents flowing across the system.  Examples of such currents are the reaction rates\cite{AG04} or the velocity of a molecular motor.\cite{AG06PRE,GG10}  At the microscopic level of description, the instantaneous currents ${\bf j}(t) = \{ j_{\gamma}(t)\}_{\gamma=1}^c$ can be defined when particles cross fictitious surfaces separating reactants from products,\cite{AG07JSP} as in reaction rate theory.  The instantaneous currents can be integrated over some time interval $t$ to get the numbers of particles having crossed the fictitious surface during that time interval: $\Delta N_{\gamma}=\int_0^t j_{\gamma}(t')\, dt'$.  As long as the time interval $t$ is finite, the currents defined as
\be
J_{\gamma} = \frac{\Delta N_{\gamma}}{t} = \frac{1}{t}\int_0^t j_{\gamma}(t')\, dt' \qquad (\gamma=1,2,...,c)
\ee
are random variables.  For given values of the different independent affinities ${\bf A}=\{A_{\gamma}\}_{\gamma=1}^c$,  the nanosystem reaches a stationary state in the long-time limit $t\to+\infty$.  This stationary state is described by a probability distribution $P_{\bf A}$.  Since the currents are fluctuating, they may take positive or negative values ${\bf J}=\{ J_{\gamma}\}_{\gamma = 1}^c$ with certain probabilities $P_{\bf A}({\bf J})$.

Now, we compare the probabilities of opposite values for the currents, $P_{\bf A}({\bf J})$ and $P_{\bf A}(-{\bf J})$.  In general, these probabilites are different but, most remarkably, the time-reversal symmetry of the underlying microscopic dynamics implies that the ratio of these probabilities has a general behavior expressed by the {\bf current fluctuation theorem}:\cite{AG07JSP}
\be
\frac{P_{\bf A}({\bf J})}{P_{\bf A}(-{\bf J})} \simeq \exp \left( {\bf A}\cdot{\bf J} \, t\right) \qquad \mbox{for}\quad t\to +\infty
\label{FTC}
\ee
This result holds for the equilibrium as well as the nonequilibrium stationary states at any value of the affinities in Markovian or semi-Markovian stochastic processes if the large-deviation properties of the process are well defined in the long-time limit.\cite{AG07JSP,AG07JSM,AG08JSM,AG09JSM}

At equilibrium where the affinities vanish, the exponential function takes the unit value and we recover the principle of detailed balancing according to which the probabilities of opposite fluctuations are equal: $P_{\bf 0}({\bf J})\simeq P_{\bf 0}(-{\bf J})$.  However, out of equilibrium when the affinities do not vanish, the ratio of probabilities typically increases or decreases exponentially in time depending on the sign of ${\bf A}\cdot{\bf J}=\sum_{\gamma=1}^c A_{\gamma} J_{\gamma}$.  Therefore, a bias grows between the probabilities of opposite fluctuations and the current fluctuations soon become more probable in one particular direction.  Directionality has thus appeared in the system.  This directionality is controlled by the affinities because the currents would flow in the opposite direction if the affinities were reversed, as expected from microreversibility.

The current fluctuation theorem has several implications.  As a consequence of Jensen's inequality, $\left\langle{\rm e}^{-X}\right\rangle \geq {\rm e}^{-\langle X\rangle}$, the thermodynamic entropy production is always non negative:
\be
\frac{1}{k_{\rm B}} \frac{d_{\rm i}S}{dt}\Big\vert_{\rm st} = {\bf A}\cdot \langle{\bf J}\rangle_{\bf A} \geq 0 
\ee
where $\langle{\bf J}\rangle_{\bf A}$ are the mean values of the currents in the stationary state  described by the probability distribution $P_{\bf A}$.  Therefore, the second law of thermodynamics is the consequence of the current fluctuation theorem.  Furthermore, this theorem allows us to generalize the Onsager reciprocity relations and the Green-Kubo formulas from the linear to the nonlinear response properties of the average currents with respect to the affinities.\cite{AG04,AG07JSM}  This generalization is the result of the validity of the current fluctuation theorem far from equilibrium.

In particular, these results apply to the reversible Brusselator model of oscillatory reactions.\cite{AG08JCP}  For fully irreversible reactions, in which the rates of the reversed reactions vanish, the corresponding affinities take infinite values so that the entropy production is also infinite, in which case the ratio of the probabilities of opposite fluctuations is consistently either zero or infinite.  The fact that the ratio (\ref{FTC}) behaves exponentially means that the reversed processes may soon become so rare that their probabilities are negligible and the system be in a far-from-equilibrium regime which could be considered as fully irreversible.

\subsection{Thermodynamic origins of dynamical order}

Another time-reversal symmetry relationship concerns the statistical properties of the histories or paths followed by a system under stroboscopic observations at some sampling time $\Delta t$.  Such observations generate a sequence of coarse-grained states:
\be
\pmb{\omega} = \omega_1\omega_2 \cdots \omega_n
\ee
corresponding to the successive times $t_j=j\, \Delta t$ with $j=1,2,...,n$.  This history or path has a certain probability $P_{\bf A}(\pmb{\omega})$ to happen if the system is in the stationary state corresponding to the affinities $\bf A$.  Because of the randomness of the molecular fluctuations, these path probabilities typically decrease exponentially as
\be
P_{\bf A}(\pmb{\omega}) = P_{\bf A}(\omega_1\omega_2 \cdots \omega_n) \sim {\rm e}^{-n \, \Delta t \, h_{\bf A}}
\ee
at a rate $h_{\bf A}$ that characterizes the temporal disorder in the process.  Such a characterization concerns stochastic processes as well as chaotic dynamical systems.\cite{GW93,G98}
In nonequilibrium stationary states, the time-reversed path
\be
\pmb{\omega}^{\rm R} = \omega_n  \cdots \omega_2\omega_1
\ee
is expected to happen with a different probability
\be
P_{\bf A}(\pmb{\omega}^{\rm R}) = P_{\bf A}(\omega_n  \cdots \omega_2\omega_1) \sim {\rm e}^{-n \, \Delta t \, h_{\bf A}^{\rm R}}
\ee
decreasing at a different rate $h_{\bf A}^{\rm R}$ now characterizing the temporal disorder of the time-reversed paths.\cite{G04JSP}  The remarkable result is that the difference between the disorders of the time-reversed and typical paths is equal to the thermodynamic entropy production:\cite{G04JSP}
\be
\frac{1}{k_{\rm B}} \frac{d_{\rm i}S}{dt}\Big\vert_{\rm st} = h_{\bf A}^{\rm R} - h_{\bf A} \geq 0 
\label{hr-h}
\ee
The second law of thermodynamics is satisfied because this difference is known in mathematics to be always non negative.\cite{CT06}  The validity of the formula (\ref{hr-h}) has been verified in experiments where the nonequilibrium constraints are imposed by fixing the currents instead of the affinities, in which case the comparison should be carried out between $P_{\bf J}(\pmb{\omega})$ and $P_{-\bf J}(\pmb{\omega}^{\rm R})$.\cite{AGCGJP07,AGCGJP08}

At equilibrium, detailed balancing holds so that every history and its time reversal are equiprobable, their temporal disorders are equal, and the entropy production vanishes.  This is no longer the case away from equilibrium where the typical paths are more probable than their time reversals.  Consequently, the time-reversal symmetry is broken at the statistical level of description in terms of the probability distribution $P_{\bf A}$ of the nonequilibrium stationary state.  In this regard, the entropy production is a measure of the time asymmetry in the temporal disorders of the typical histories and their time reversals.  As a corollary of the second law, we thus have the {\bf theorem of nonequilibrium temporal ordering}:\cite{G07CRP}

{\it In nonequilibrium stationary states, the typical histories are more ordered in time than their corresponding time reversals in the sense that $h_{\bf A} < h_{\bf A}^{\rm R}$.}

This temporal ordering is possible out of equilibrium at the expense of the increase of phase-space disorder so that there is no contradiction with Boltzmann's interpretation of the second law.  The result established by this theorem is that nonequilibrium processes can generate dynamical order, which is a key feature of biological phenomena.  In particular, the nonequilibrium ordering mechanism can generate oscillatory behavior in surface reactions or information-rich sequences in copolymers.

\section{Heterogeneous catalytic reactions in high electric fields}
\label{Catalysis}

Dynamical order in the form of nonequilibrium patterns or oscillations can arise in heterogeneous catalysis at the nanoscale on metallic tips under FEM or FIM conditions.\cite{vTGN92,GLRBE94,EBGKWB94,VK96,KV08}

\subsection{Surface conditions in FEM and FIM}

Since solid metals are crystalline and the radius of curvature of typical tips may reach 10-30 nm, the surface is  multifaceted, which introduces non-uniformities as shown by Gerhard Ertl and coworkers.\cite{GLRBE94}  Moreover, the surface is subjected to high electric fields of about 10 V/nm, which have influence on the surface reactions.\cite{KV08}  The electrostatic edge effect tends to concentrate the electric field on the sharpest structures of the metallic needle.  This is the case at the edges of the crystalline facets, which provides the atomic resolution of cryogenic FIM.\cite{MT69}  At a larger scale, this is also the case at the apex of the tip where the radius of curvature of the average surface is the smallest.  If the average shape of the tip is a paraboloid, the electric field varies as
\be
F = \frac{F_0}{\sqrt{1+\frac{r^2}{R^2}}}
\label{field}
\ee
as a function of the radial distance $r$ from the axis of cylindrical symmetry of the paraboloid.  The quantity $R$ denotes the radius of curvature at the apex where the electric field reaches its maximum value $F_0$.  

This high electric field has several effects, which create the conditions of a nanoreactor localized near the apex of the needle under electric tension.  On the one hand, the electric field polarizes the molecules in the gaseous mixture around the needle.  Consequently, the partial pressures  increase according to
\be
P(F) \simeq P(0) \, \exp\frac{\alpha F^2}{2k_{\rm B}T}
\label{pressure}
\ee
where $\alpha$ is an effective polarizability of the molecules of a given species and $k_{\rm B}$ Boltzmann's constant.\cite{VKGK06}  On the other hand, the electric field modifies the activation energies of the different processes taking place on the surface:
\be
E_{\rm x}(F) = E_{\rm x}(0) - d _{\rm x} F - \frac{\alpha _{\rm x}}{2} F^2  + \cdots
\ee
The dependence of the activation energies is in general a nonlinear function of the electric field $F$.  At sufficiently low values of the electric field, such power expansions define the coefficients $d _{\rm x}$, $\alpha _{\rm x}$,..., which are associated with the transition state in analogy to the situation in the stable states.  

Since the surface is multifaceted, its properties also depend on the orientation of every crystalline plane where the reactions proceed.  This dependence can be expressed in terms of the corresponding Miller indices $(h,k,l)$ or, equivalently, the unit vector normal to the plane:
\be
{\bf n} = (n_x,n_y,n_z) = \frac{(h,k,l)}{\sqrt{h^2+k^2+l^2}}
\label{n_vector}
\ee
This anisotropy concerns, in particular, the activation energies, which can be expanded in kubic harmonics as\cite{MGVK09a,MGVK09b,MGVK10}
\be
E_{\rm x}({\bf n}) = E_{\rm x}^{(0)} + E_{\rm x}^{(4)} (n_x^4+n_y^4+n_z^4) + E_{\rm x}^{(6)} (n_x^2 \, n_y^2 \, n_z^2) + \cdots
\label{kubic}
\ee
for face-centered cubic crystals such as rhodium or platinum.  The coefficients of this expansion can be fitted to data collected for different orientations.  The knowledge of any activation energy for the three main crystalline planes $(001)$, $(011)$, and $(111)$ determines the three first coefficients of the expansion (\ref{kubic}).  Finer dependences of the activation energy can be included with experimental data on more crystalline planes.  These dependences on the various facets composing the tip are crucial to understand the anisotropy of the surface reactions and the nanopatterns observed under FEM or FIM conditions.
 
 These conditions are significantly different from those prevailing on flat crystalline surfaces extending over distances of several hundreds or thousands of micrometers.  Such flat surfaces have uniform properties so that the transport mechanisms are diffusive and the patterns observed on flat crystalline surfaces emerge as the result of standard reaction-diffusion processes in uniform media.  The wavelengths of such reaction-diffusion patterns are determined by the diffusion coefficients $D$ and the reaction constants $k_{\rm rxn}$.  They are of the order of $L \sim \sqrt{D/k_{\rm rxn}} \sim 100\, \mu{\rm m}$, which is much larger than the size $R\simeq 20$ nm of a FIM tip.  Therefore, the nanopatterns observed under FEM or FIM conditions are not standard reaction-diffusion patterns and their understanding requires to take into account the non-uniform and anisotropic effects of the electric field and the underlying crystal. On non-uniform surfaces, the transport of adsorbates is not only driven by the gradients of coverages, but also by the surface gradients of desorption energy and electric field.\cite{ME95,HME98,DMHGKMI04,GMG11}

\subsection{Adsorption-desorption kinetics}

For diatomic molecules A$_2$ such as dihydrogen, adsorption is dissociative.
Therefore, the coverage $\theta_{\rm A}$ of the surface by the atomic species increases at the rate:
\be
\frac{\partial\theta_{\rm A}}{\partial t}\Big\vert_{\rm ads} = 2\, k_{\rm a} \, P_{{\rm A}_2}  \, (1-\theta_{\rm A})^2 
\qquad\mbox{with}\qquad
k_{\rm a} = \frac{S^0 a_{\rm s}}{\sqrt{2\pi m_{{\rm A}_2} k_{\rm B} T}}
\ee
where the pressure $P_{{\rm A}_2}$ is given by Eq.~(\ref{pressure}), $S^0$ is the initial sticking coefficient at zero coverage, and $a_{\rm s}$ is some reference area.
Accordingly, desorption is dissociative and proceeds at the thermally activated rate
\be
\frac{\partial\theta_{\rm A}}{\partial t}\Big\vert_{\rm des} = - 2\, k_{\rm d} \, \theta_{\rm A}^2
\qquad\mbox{with}\qquad
k_{\rm d} =k_{\rm d}^0\, {\rm e}^{-\beta E_{\rm d}({\bf n},F,\theta_{\rm A})}
\ee
and $\beta=(k_{\rm B}T)^{-1}$.  The desorption energy $E_{\rm d}$ depends on the electric field (\ref{field}), the local crystalline orientation (\ref{n_vector}), as well as the coverage itself if lateral interactions play a role in desorption.

Experimental data are available for the adsorption and desorption of hydrogen on rhodium
\be
\mbox{H$_2$ (gas)} \; +\;  2\; \emptyset \; \mbox{(ad)}
\ \underset{k_{\rm dH}}{\overset{k_{\rm aH}}{\rightleftharpoons}}
\ 2\;  \mbox{H (ad)}
\label{H-ads-des}
\ee
Its desorption energy takes values below one electron-Volt so that the mean hydrogen coverage $\theta_{\rm H}$ is low above 400\,K where bistability and oscillations are observed.\cite{MGVK09a,MGVK09b,MGVK10}

\subsection{Surface oxides of rhodium}

The behavior is more complex for oxygen on rhodium.
On the one hand, the dissociative adsorption of oxygen involves a precursor state, which explains the dependence of the sticking coefficient on the oxygen coverage.\cite{SLWC90,SS93}

On the other hand, oxygen forms surface oxide trilayers on rhodium: O(ad)-Rh-O(sub).  Their properties have been systematically investigated, in particular, with DFT calculations.\cite{GRS02,GMBLKKSVYTQA04,AELCHRC04,KKSLGMBYAMV04,GMBALKHSVKKKSD05,DAECDKMKDK06,LMAKSV06,M10}  By their stoichiometry RhO$_2$, the surface oxides differ from the bulk oxide Rh$_2$O$_3$. Oxygen vacancies can thus exist in either the outer or the inner oxygen layer.  Accordingly, the structure of a partially formed surface oxide can be characterized in terms of the occupancies by oxygen atoms of the adsorption and subsurface sites, $\theta_{\rm O}$ and $\theta_{\rm s}$ respectively.

These different features of the interaction of oxygen with rhodium are described by the following kinetic scheme:\cite{MGVK09a,MGVK09b,MGVK10}
\bea
&&\mbox{adsorption and desorption:} \qquad \mbox{O$_2$ (gas)} \; + \; \mbox{surface} \; 
\underset{\tilde k_{\rm dO}}{\overset{\tilde k_{\rm aO}}{\rightleftharpoons}} \ \mbox{O$_2$ (pre)}  \; + \; \mbox{surface} \; 
\nonumber \\ 
&&\mbox{dissociation and recombination:} \qquad
\mbox{O$_2$ (pre)} \; +\;  2\; \emptyset \; \mbox{(ad)}
\ \underset{k_{\rm dO}}{\overset{k_{\rm aO}}{\rightleftharpoons}}
  \ 2\;  \mbox{O (ad)}  \nonumber \\ 
  &&\mbox{oxidation and reduction of Rh:} \qquad
\mbox{O (ad)} \; + \; \emptyset\; \mbox{(sub)}
\ \underset{k_{\rm red}}{\overset{k_{\rm ox}}{\rightleftharpoons}}
  \  \emptyset\; \mbox{(ad)} \; + \;  \mbox{O (sub)} 
\nonumber \\ &&
  \label{O-Rh}
\eea
The surface oxide trilayer tends to inhibit oxygen adsorption, which is taken into account by the precursor constant:\cite{MGVK09a,MGVK09b,MGVK10}
\be
K=\frac{k_{\rm aO}}{\tilde k_{\rm dO}}=K^0\, {\rm e}^{-\beta(E_K+A_{K}^{\rm O}\theta_{\rm O}+A_{K}^{\rm s}\theta_{\rm s})}
\label{Kpre}
\ee
with parameters fitted to data on the sticking coefficient.\cite{SLWC90,SS93}  Besides, the desorption rate constants on the three main surface orientations are fitted to temperature-programmed desorption spectra.\cite{MGVK09a,MGVK09b,MGVK10}

For rhodium in equilibrium with gaseous dioxygen, the adsorbate and subsurface occupancies satisfy
\be
\frac{\theta_{\rm O}}{1-\theta_{\rm O}} = \frac{k_{\rm red}}{k_{\rm ox}} \frac{\theta_{\rm s}}{1-\theta_{\rm s}} = \sqrt{\frac{k_{\rm aO}}{k_{\rm dO}}\frac{\tilde k_{\rm aO}}{\tilde k_{\rm dO}} P_{{\rm O}_2}}
\ee
A phase transition between a metallic surface with adsorbed oxygen and a surface oxide trilayer with possible vacancies occurs at about $\theta_{\rm s}\simeq 0.5$.
This condition allows us to determine the ratio of the oxidation and reduction rates of a rhodium layer using the results of DFT calculations on the three main surface orientations.\cite{MGVK09a,MGVK09b,MGVK10}  

Moreover, the dependence of the activation energies on the electric field has been evaluated using DFT calculations,\cite{MGMVK08} which shows that a positive electric field tends to promote the oxidation of rhodium.  

\subsection{The H$_2$-O$_2$/Rh system}

Water formation is catalyzed on rhodium in contact with a gaseous mixture of dihydrogen and dioxygen.  If water vapor is not supplied in the mixture, its partial pressure is vanishing.  Under these conditions, the corresponding affinity (\ref{affinity}) is infinite and the overall reaction proceeds in a fully irreversible regime.

For the reaction of water formation on rhodium
\be
2 \; \mbox{H (ad)} \; + \; \mbox{O (ad)} 
\; \ {\overset{k_{\rm r}}{\rightarrow}}
\;  \   3\; \emptyset\; \mbox{(ad)} \; + \; \mbox{H}_2\mbox{O (gas)} 
\label{react}
\ee
the rate constant is taken as
\be
k_{\rm r} = k_{\rm r}^0 \, {\rm e}^{-\beta (E_{\rm r}-d_{\rm r}F+A_{\rm r}^{\rm H}\theta_{\rm H}+A_{\rm r}^{\rm O}\theta_{\rm O})}
\label{kr}
\ee
with coefficients $A_{\rm r}^{\rm H}$ and  $A_{\rm r}^{\rm O}$ expressing
the change of the reaction rate with the hydrogen and oxygen coverages, respectively.\cite{MGVK09a,MGVK09b,MGVK10}  

\subsubsection{Kinetic equations}

Combining together the different processes involving hydrogen and oxygen, the kinetic equations are given by\cite{MGVK09a,MGVK09b,MGVK10}
\bea
\frac{\partial\theta_{\rm H}}{\partial t} &=& 2\, k_{\rm aH} \, P_{{\rm H}_2}  \, 
\theta_{\emptyset}^2 - 2\, k_{\rm dH} \, \theta_{\rm H}^2 - 2\, k_{\rm r} \, \theta_{\rm H} \, \theta_{\rm O} -{\rm div}\, {\bf J}_{\rm H} \label{eq1}\\
\frac{\partial\theta_{\rm O}}{\partial t} &=& \frac{2}{1+ K  \, \theta_{\emptyset}^2} \, \left( \tilde k_{\rm aO} \, K \, P_{{\rm O}_2} \, \theta_{\emptyset}^2 
- k_{\rm dO} \, \theta_{\rm O}^2\right) - \, k_{\rm ox} \, \theta_{\rm O} \, \left( 1 - \theta_{\rm s}\right) + k_{\rm red} \, \theta_{\rm s} \, \theta_{\emptyset} - k_{\rm r} \, \theta_{\rm H} \, \theta_{\rm O} \label{eq2} \\
\frac{\partial\theta_{\rm s}}{\partial t} &=& k_{\rm ox}\, \theta_{\rm O} \, \left( 1 - \theta_{\rm s}\right)
- k_{\rm red} \, \theta_{\rm s} \, \theta_{\emptyset} \label{eq3}
\eea
with the coverage of empty sites $\theta_{\emptyset} = 1-\theta_{\rm H}-\theta_{\rm O}$ and $K=k_{\rm aO}/\tilde k_{\rm dO}$.  The current density of hydrogen takes the form
\be
{\bf J}_{\rm H} = - D_{\rm H} \left[ (1- \theta_{\rm O}) \pmb{\nabla} \theta_{\rm H} + \theta_{\rm H}
\pmb{\nabla} \theta_{\rm O} + \theta_{\rm H} (1-\theta_{\rm H}-\theta_{\rm O}) \frac{\pmb{\nabla} U_{\rm H}}{k_{\rm B} T} \right]
\ee
with the effective energy potential
\be
U_{\rm H}({\bf r})=-\frac{1}{2} \left[ E_{\rm dH}({\bf r}) - d_{\rm dH} F({\bf r}) + \frac{1}{2} \alpha_{{\rm H}_2} F({\bf r})^2 \right] + {\rm cst}
\label{UH}
\ee

The mobility of oxygen is negligible.  In contrast, the mobility of hydrogen is very high with a typical diffusion time $t_{\rm diff}=R^2/D_{\rm H}\simeq 10^{-7}$\,s, which is many orders of magnitude shorter than the time scales of the other kinetic processes at 500\,K.  Therefore, the coverage of atomic hydrogen quickly reaches the quasi-equilibrium distribution
\be
\theta_{\rm H}({\bf r},t) = \frac{1-\theta_{\rm O}({\bf r},t)}{\displaystyle 1+{\rm e}^{\beta\left[ U_{\rm H}({\bf r})-\mu_{\rm H}(t)\right]}}
\label{H.quasieq.distrib}
\ee
such that the hydrogen current density is vanishing: ${\bf J}_{\rm H}=0$.
This coverage varies in space because the effective potential (\ref{UH}) depends on the electric field~(\ref{field}) and on the normal unit vector (\ref{n_vector}), which is uniquely determined by the position $\bf r$ on the surface.  Therefore, the hydrogen coverage already forms nanopatterns, which reflect the anisotropy of the underlying crystal and the electric field variation.  The chemical potential $\mu_{\rm H}$ is uniform on the surface but slowly
varies in time because the population of hydrogen atoms 
on the surface changes due to adsorption, desorption and the reaction of hydrogen with oxygen.
Using multiscale analysis, it has been possible to show that this nonequilibrium chemical potential evolves in time according to
\bea
\frac{d\mu_{\rm H}}{dt} &=& k_{\rm B} T \, \frac{\sum_{\rm facets} w \left[ (1-\theta_{\rm O})\partial_t \theta_{\rm H}+\theta_{\rm H}\partial_t \theta_{\rm O}\right]}{\sum_{\rm facets} w\, \theta_{\rm H}(1-\theta_{\rm H}-\theta_{\rm O})} 
\label{eq4}
\eea
where $w=(1-\theta_{\rm O})^{-1}$, while the time derivatives $\partial_t \theta_{\rm H}$ and $\partial_t \theta_{\rm O}$ are given by the kinetic Eqs. (\ref{eq1}) and (\ref{eq2}).\cite{GMG11} 

This kinetic model reproduces very well the nonlinear dynamics of the system and the spatial dependence of the observed nanopatterns.

\subsubsection{Bistability}

\begin{figure}[h]
\centerline{\includegraphics[width=5cm]{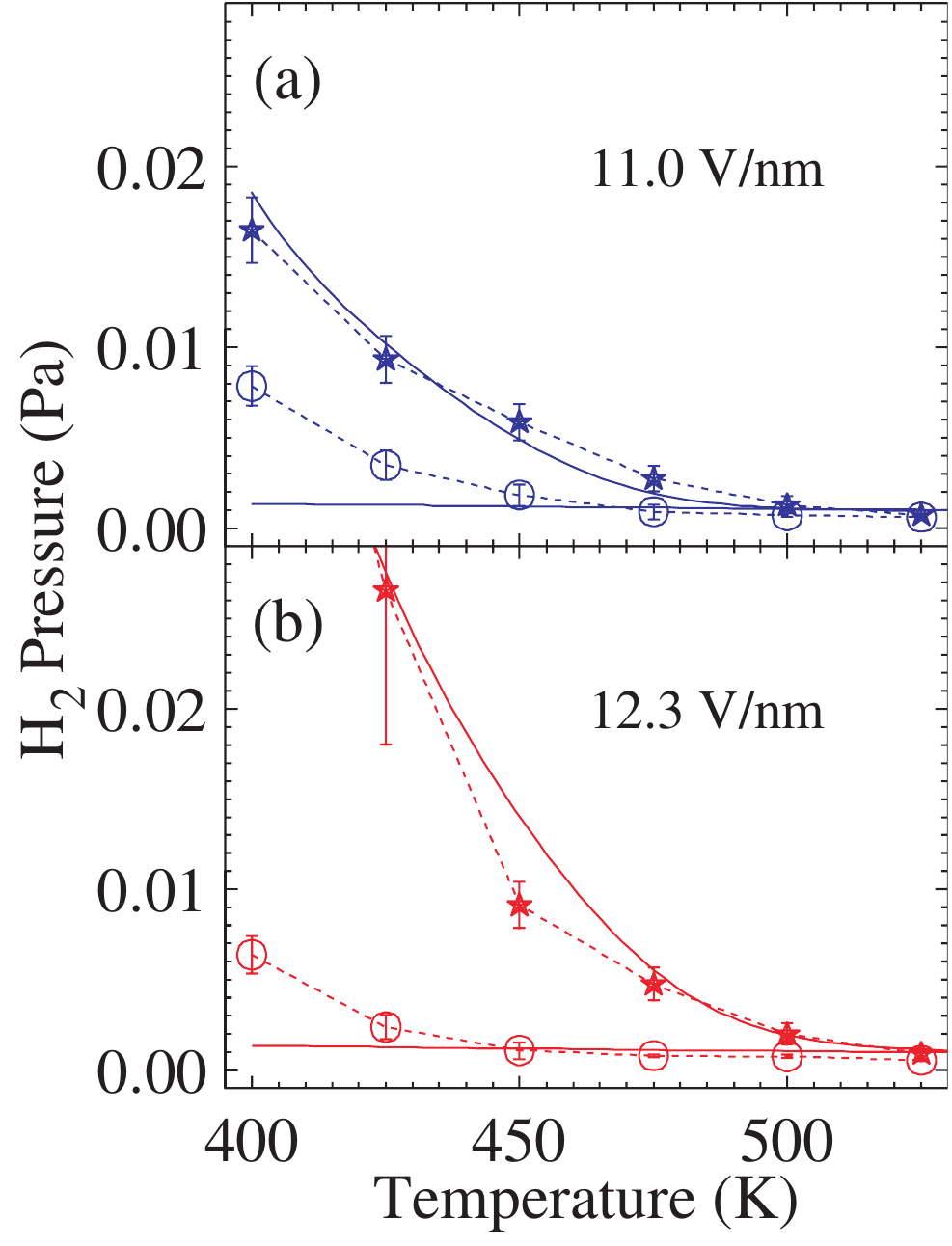}}
\caption{The bistability diagram showing hysteresis for $P_{\mathrm{O}{_2}}=5\times10^{-4}$ Pa during catalytic water formation on rhodium at (a) 11.0 V/nm and (b) 12.3 V/nm.  The circles and 
stars indicate the experimental pressures for which the structural transformation occurred when 
decreasing and increasing the hydrogen pressure, respectively.  The area in between marks the coexistence region of bistability.  The full lines are the corresponding results for the kinetic model ruled by Eqs.~(\ref{eq1})-(\ref{eq4}).\cite{MGVK09a,MGVK09b}} 
\label{fig1}
\end{figure}

Bistability manifests itself under variation of hydrogen pressure for fixed oxygen pressure, as shown in Fig.~\ref{fig1} as a function of temperature.\cite{MGVK09a,MGVK09b,MGVK10}  At low (resp. high) hydrogen pressure, the surface is covered with oxygen (resp. hydrogen).  A domain of hysteresis appears in between where the two states coexist.  The bistability domain depends on the applied electric field.  The higher the field, the broader the coexisting region in the bifurcation diagram of Fig.~\ref{fig1}.  As temperature increases, so does the water formation rate, leading to the reduction of the oxide and the coverage of the tip by hydrogen.  Ultimately, the bistability domain disappears above 550\,K.

\subsubsection{Oscillations}

\begin{figure}[h]
\centerline{\includegraphics[width=10cm]{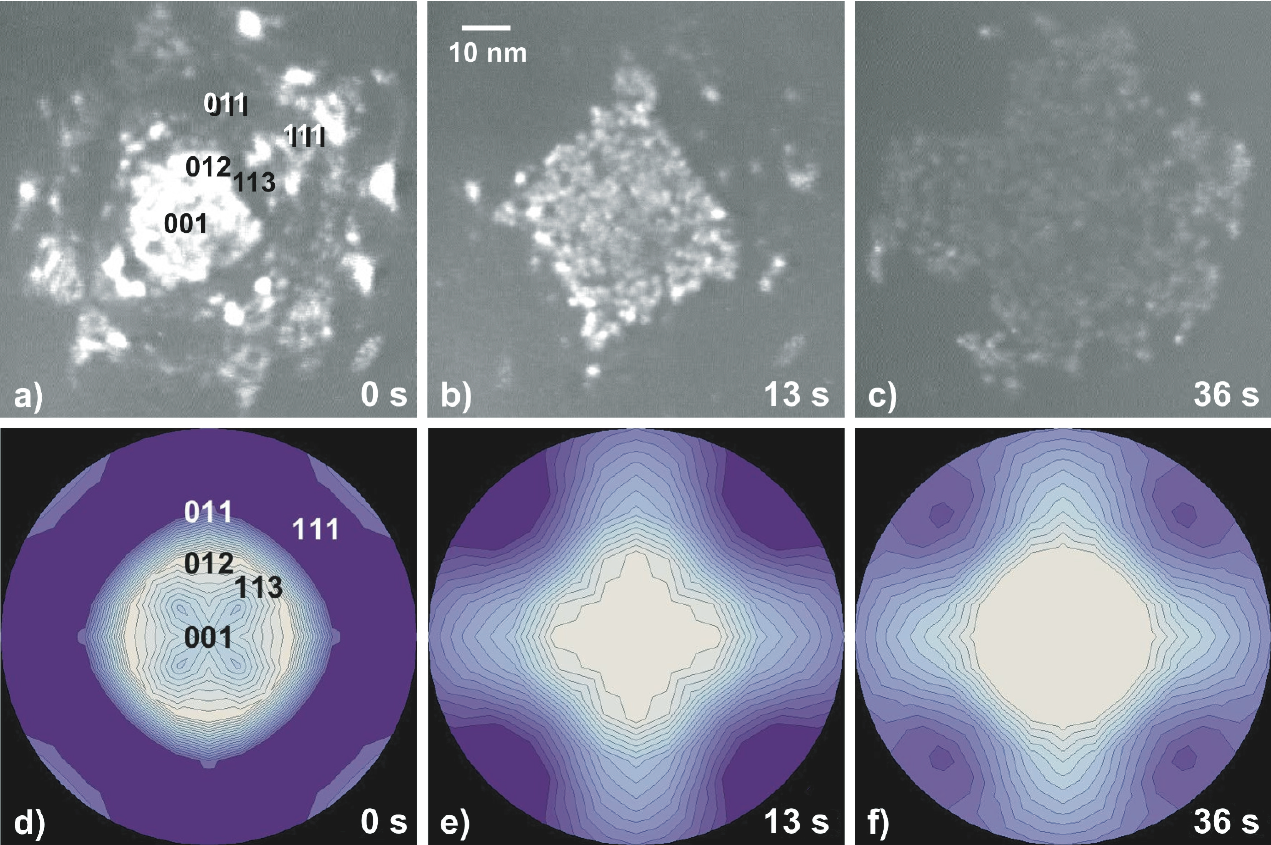}}
\caption{Series of FIM micrographs covering the complete oscillatory cycle as well as the corresponding time evolution of the subsurface oxygen distribution on a logarithmic scale as obtained within the kinetic model.\cite{MGVK09a,MGVK09b}  The temperature, electric field and partial pressures 
of oxygen are $T = 550$ K, $F_0= 12$ V/nm, $P_{\mathrm{O}{_2}}=2\times10^{-3}$ Pa, 
respectively.  On the other hand, the hydrogen pressure is 
$P_{\mathrm{H}{_2}}=2\times10^{-3}$ Pa in the FIM experiments 
and $4\times10^{-3}$ Pa in the simulation of the kinetic model.  
For the subsurface site occupation, the white areas indicate a high 
site occupation value while the dark areas indicate a low site occupation value.} 
\label{fig2}
\end{figure}

\begin{figure}[h]
\centerline{\includegraphics[width=6cm]{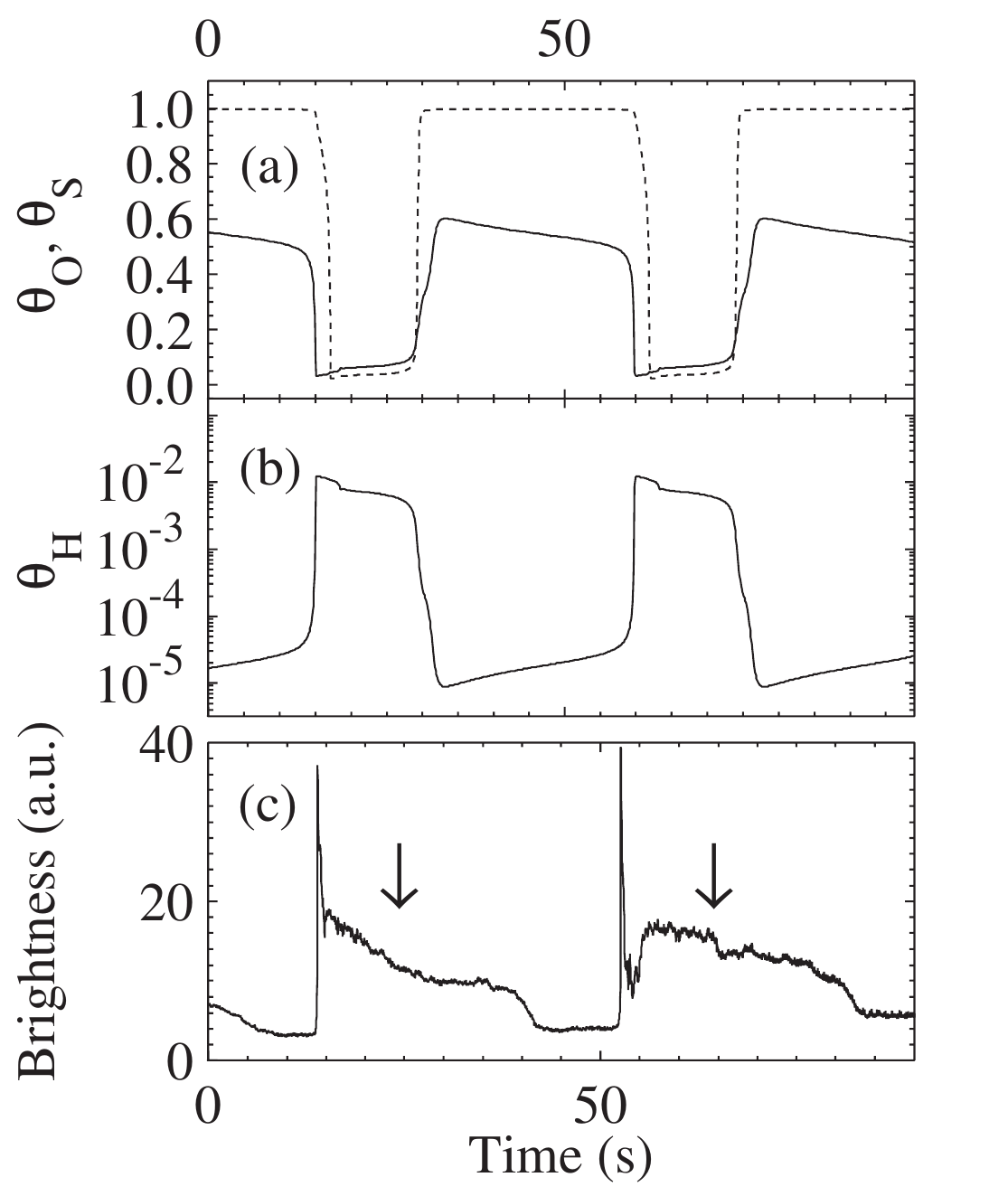}}
\caption{(a) The time evolution of the oxygen coverage (solid line) and the oxygen subsurface occupation (dashed line) at the (001) plane in the kinetic model in the oscillatory regime.  (b) Corresponding oscillations of the hydrogen coverage at the (001) plane.  (c)~Experimental total brightness during the oscillations.\cite{MGVK09a}  The conditions are the same as in Fig.~\ref{fig2}.  The arrows indicate the transition from a metallic rhodium field emitter tip to one that is invaded by subsurface oxygen.} 
\label{fig3}
\end{figure}

The kinetic model also explains the oscillatory behavior observed in this system in FIM experiments (see Figs.~\ref{fig2} and~\ref{fig3}).\cite{MGVK09a,MGVK09b,MGVK10}  The period of oscillations is about 40\,s.  In the model, this period is mainly determined by the rate constants of rhodium oxidation and reduction when oxygen reacts with the first rhodium layer.  The feedback mechanism at the origin of the oscillations involves, in particular, the formation of surface oxide and its inhibition of further oxygen adsorption, as taken into account with the rate constant (\ref{Kpre}).

Starting from a quasi metallic surface in Figs.~\ref{fig2}a~\&~\ref{fig2}d, 
an oxide layer invades the topmost plane and grows along the \{011\} facets forming a nanometric cross-like structure seen in Figs.~\ref{fig2}b~\&~\ref{fig2}e.  The oxide front spreads to cover finally the whole visible surface area in Figs.~\ref{fig2}c~\&~\ref{fig2}f.  This is associated with a decrease of the overall brightness in~Fig.~\ref{fig3}c.  The oscillation cycle is closed by a sudden reduction of the surface oxide from the outskirts towards the top, with a considerable increase of the brightness.  During the cycle shown in Fig.~\ref{fig3}, rhodium is alternatively covered by the surface oxide when $\theta_{\rm s}\simeq1$ and $\theta_{\rm O}\simeq 0.5$ and, thereafter, by hydrogen at a low coverage since the temperature is 550\,K.

\subsection{Self-organization at the nanoscale}

In summary, the nonlinear dynamics and the patterns observed in field emission microscopy are determined by the tip geometry, the anisotropy from the underlying crystal, and the electric field.  The anisotropy can be described by giving to the energy barriers their dependence on the crystalline orientation for the many facets composing the tip of the field emission microscope, in particular, with the systematic expansions (\ref{kubic}) into kubic harmonics for cubic crystals.
Another important feature is the ultrafast mobility of hydrogen, which remains in the quasi-equilibrium distribution (\ref{H.quasieq.distrib}) slowly evolving because of the other kinetic processes.  The kinetic model is built on the basis of experimental data about adsorption and desorption of hydrogen and oxygen as revealed by temperature-programmed desorption spectra and the recent studies on the RhO$_2$ rhodium surface oxides.\cite{SLWC90,SS93,GRS02,GMBLKKSVYTQA04,AELCHRC04,KKSLGMBYAMV04,GMBALKHSVKKKSD05,DAECDKMKDK06,LMAKSV06,M10}  By taking into account all these different aspects, the kinetic model provides a comprehensive understanding of the bistability, the oscillations, and the nanopatterns observed in FIM experiments.\cite{MGVK09a,MGVK09b,MGVK10}

Chemical nanoclocks have been observed under field emission microscopy conditions in several reactions besides water formation on rhodium.\cite{EBGKWB94,VK96,MGDBVK10}  Surprisingly, rhythmic behavior is possible at the nanoscale of 10-30 nm in the population dynamics of the different species reacting on the surface.  Indeed, this area may contain up to about ten thousand adsorbates, which is already much larger than the minimum size of a few hundred molecules required to sustain correlated oscillations.\cite{G02} This behavior is an example of dynamical order, which can manifest itself out of equilibrium as a corollary of the second law.

\section{Copolymerization processes}
\label{Copolymers}

\subsection{Information processing at the molecular scale}

If the history of a nonequilibrium system can be recorded on a spatial support of information, the theorem of nonequilibrium temporal ordering\cite{G07CRP} suggests that dynamical order may generate regular information sequences, which is not possible at equilibrium.

At the molecular scale, natural supports of information are given by random copolymers where information is coded in the covalent bonds.  This is the idea of Schr\"odinger's aperiodic crystal.\cite{S44}    Random copolymers exist in chemical and biological systems.  Examples are styrene-butadiene rubber, proteins, RNA, and DNA, this latter playing the role of information support in biology.
Accordingly, dynamical aspects of information are involved in copolymerization processes where fundamental connections with nonequilibrium thermodynamics have been recently discovered.\cite{AG08,J08,AG09}

\subsection{Thermodynamics of free copolymerization}

The stochastic growth of a single copolymer proceeds by attachment and detachment of monomers $\{m\}$ continuously supplied by the surrounding solution, which is assumed to be sufficiently large to play the role of a reservoir where the concentrations of the monomers are kept constant:
\be
\omega = m_1m_2\cdots m_{l}  \ \underset{-m_{l+1}}{\overset{+m_{l+1}}{\rightleftharpoons}}
  \  \omega' = m_1m_2\cdots m_lm_{l+1}
\ee
The probability $P_t(\omega)$ to find the monomer sequence $\omega$ of length $l=\vert\omega\vert$ at the time $t$ is ruled by the master equation
\be
\frac{d P_t(\omega)}{dt} = \sum_{\omega'} \left[
P_t(\omega') \, W(\omega'\vert\omega)
- P_t(\omega) \, W(\omega\vert\omega')\right] 
\label{master.eq}
\ee
where the coefficients $W(\omega\vert\omega')$ denote the rates of the transitions
$\omega\to\omega'$.  If attachment and detachment processes are slower than the equilibration of the chain with its environment, the transition rates satisfy the conditions of local detailed balancing
\be
\frac{W(\omega\vert\omega')}{W(\omega'\vert\omega)} = \exp
\frac{G(\omega)-G(\omega')}{k_{\rm B} T} \, ,
\label{ratio}
\ee
in terms of the Gibbs free energy $G(\omega)$ of a single copolymer chain
$\omega$ in the solution at the temperature $T$.  The enthalpy $H(\omega)$ and the entropy $S(\omega)$ of the copolymer chain $\omega$ can similarly be defined and they are related together by $G(\omega)=H(\omega)-TS(\omega)$.  At a given time $t$, the system may be in different sequences and different configurations so that the total entropy has two contributions:
\be
S_t = \sum_{\omega} P_t(\omega) S(\omega) - k_{\rm B}
\sum_{\omega} P_t(\omega) \ln P_t(\omega) 
\label{entropy}
\ee
The first one is due to the statistical average of the phase-space disorder $S(\omega)$ of the individual copolymer chains $\omega$ and the second is due to the probability distribution itself among the different possible sequences $\omega$ existing at the current time $t$.

In the regime of steady growth,\cite{CF63JPS,CF63JCP} this probability is supposed to be factorized as
\be
P_t(\omega) = \mu_l(\omega) \times p_t(l)
\ee
into a stationary statistical distribution $\mu_l(\omega)$ describing the arbitrarily long sequence which is left behind, multiplied by the time-dependent probability $p_t(l)$ of the length $l$ selected in the sequence.  In this regime, the mean growth velocity is constant and given by $v = d\langle l\rangle_t/dt$ where $\langle l\rangle_t=\sum_l p_t(l)\times l$ is the mean length of the copolymer at time $t$.  Mean values per monomer can be defined as
\be
x = \lim_{l\to\infty} \frac{1}{l} \sum_{\omega} \mu_l(\omega) \,X(\omega)
\ee
for entropy $s$,  enthalpy $h$, and Gibbs free energy $g=h-Ts$.  

\begin{figure}
\centerline{\includegraphics[width=4cm]{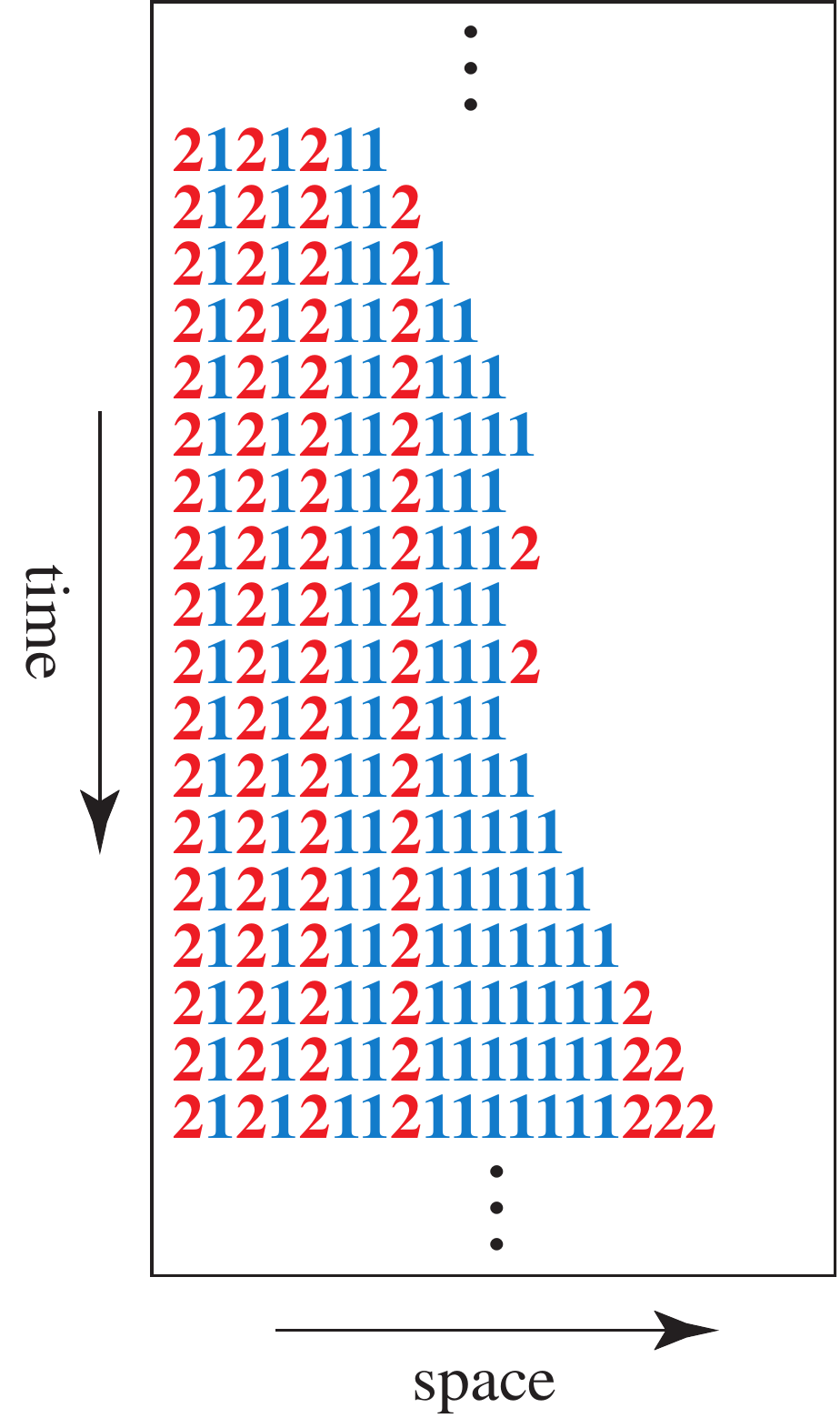}}
\caption{Space-time plot of the stochastic growth of a copolymer composed of two monomers, as simulated by Gillespie's algorithm with the parameter values:  $k_{+1}=1$, $k_{-1}=10^{-3}$, $[1]=10^{-3}$, $k_{+2}=2$, $k_{-2}=2\times 10^{-3}$, and $[2]=5\times 10^{-4}$, in the growth regime by entropic effect.\cite{AG09}  Under these conditions, the fraction of monomers $1$ is $p=0.618$, the growth velocity $v=6.17\times 10^{-4}$ (i.e., $0.18$ monomer per reactive event), the free-energy driving force $\varepsilon=-g/(k_{\rm B}T)=-0.265$, the Shannon disorder $D=0.665$, the affinity $A=\varepsilon+D=0.400$, and the entropy production $\frac{d_{\rm i}S}{dt}=Av=2.47\times 10^{-4}$  in units where $k_{\rm B}=1$.}
\label{fig4}
\end{figure}

In these circumstances, the total entropy (\ref{entropy}) can be calculated and shown to vary in time as
\be
\frac{dS_t}{dt} = \frac{d_{\rm e}S}{dt} +\frac{d_{\rm i}S}{dt}
\ee
due to the entropy exchange between the copolymer and its surrounding:
\be
\frac{d_{\rm e}S}{dt} = \frac{h}{T} \, v
\ee
and the entropy production:
\be
\frac{d_{\rm i}S}{dt} = k_{\rm B} \, A \, v \geq 0
\label{entr-prod-copolym}
\ee
which is always non negative according to the second law of thermodynamics.
The entropy production is expressed in terms of the affinity\cite{AG08}
\be
A = - \frac{g}{k_{\rm B}T} + D(\mbox{polymer})
\label{Aff}
\ee
which involves, on the one hand, the free energy per monomer $g$ and, on the other hand, the Shannon disorder per monomer in the sequence composing the copolymer:
\be
D(\mbox{polymer}) = \lim_{l\to\infty} -\frac{1}{l} \sum_{\omega} \mu_l(\omega) \,\ln \mu_l(\omega)\geq 0
\label{disorder}
\ee

\begin{figure}
\centerline{\includegraphics[width=8cm]{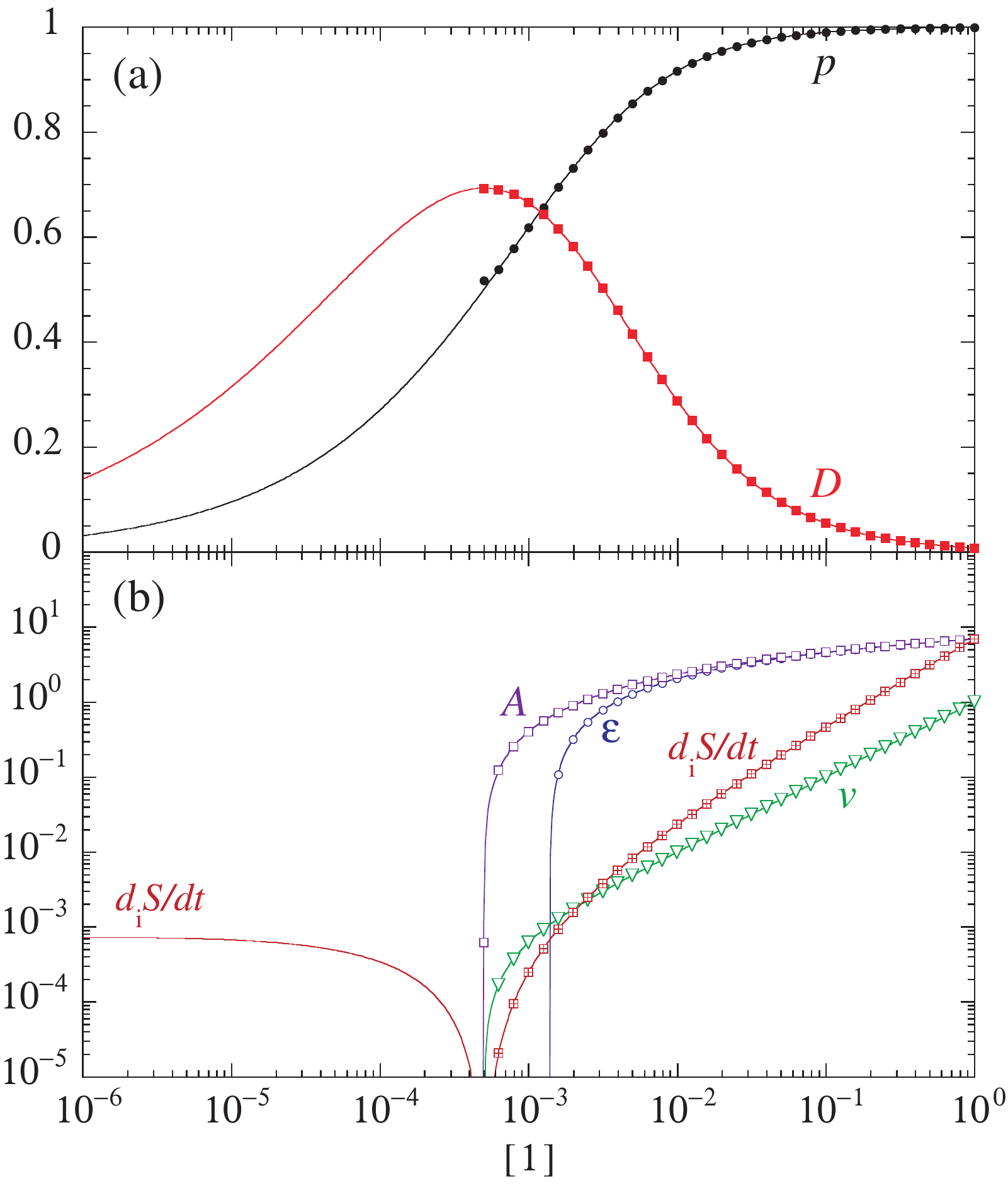}}
\caption{Comparison between simulation (dots) and theory (lines) for the growth of a copolymer composed of two monomers with the parameter values: $k_{+1}=1$, $k_{-1}=10^{-3}$, $k_{+2}=2$, $k_{-2}=2\times 10^{-3}$, and $[2]=5\times 10^{-4}$.\cite{AG09}  Several characteristic quantities are depicted versus the concentration $[1]$ of monomers~$1$: (a) the fraction $p$ of monomers $1$ in the copolymer (circles) and the Shannon disorder $D$ (squares); (b) the growth velocity $v$ (triangles), the free-energy driving force $\varepsilon$ (open circles), the affinity $A=\varepsilon+D$ (open squares), and the entropy production $\frac{d_{\rm i}S}{dt}=Av$ (crossed squares).  The entropy production vanishes at the equilibrium concentration $[1]_{\rm eq}=5\times 10^{-4}$ together with the velocity and the affinity.  However, the free-energy driving force vanishes at the larger concentration $[1]_0=1.30453\times 10^{-3}$. The regime of growth by entropy effect exists between these two values of the concentration.  The velocity and the affinity are positive for $[1]>[1]_{\rm eq}=5\times 10^{-4}$ and negative below (not shown).  The free-energy driving force is positive for $[1]> [1]_0=1.30453\times 10^{-3}$ and negative below (not shown).}
\label{fig5}
\end{figure}

A prediction of this result is that a copolymer can grow by the entropic effect of disorder $D>0$ in an adverse free-energy landscape as long as the affinity (\ref{Aff}) is positive.\cite{AG08,J08}  This is illustrated in Figs.~\ref{fig4}~and~\ref{fig5} for free copolymerization with two monomers.  The concentration $[2]$ of the second monomer is kept fixed while the concentration $[1]$ of the first one is varied.  The growth velocity, as well as the affinity (\ref{Aff}) and the entropy production (\ref{entr-prod-copolym}), all vanish at equilibrium for $[1]=[1]_{\rm eq}$, which does not coincide with the concentration $[1]=[1]_0$ where the free-energy driving force $\varepsilon=-g/(k_{\rm B}T)$ is vanishing.  Therefore, the growth is possible for intermediate values of the concentration $[1]_{\rm eq} < [1] < [1]_0$ in the entropic growth regime, preceding the growth regime driven by free energy when $\varepsilon >0$ for $[1] > [1]_0$.  At equilibrium, the Shannon disorder (\ref{disorder}) reaches its maximum value $D=\ln 2$ and decreases away from equilibrium, as seen in Fig.~\ref{fig5}a.

\subsection{Thermodynamics of copolymerization with a template}

A similar result holds for copolymerizations with a template.  If the sequence of the template $\alpha$ is characterized by the statistical distribution $\nu_l(\alpha)$, the Shannon conditional disorder of the copy $\omega$ with respect to the template is defined as\cite{AG08}
\be
D(\mbox{copy}\vert\mbox{template}) = \lim_{l\to\infty} -\frac{1}{l} \sum_{\alpha,\omega} \nu_l(\alpha) \, \mu_l(\omega\vert\alpha) \,\ln \mu_l(\omega\vert\alpha)\geq 0
\label{cond-disorder}
\ee
and the mutual information between the copy and the template as\cite{CT06}
\be
I(\mbox{copy},\mbox{template}) = D(\mbox{copy}) - D(\mbox{copy}\vert\mbox{template}) \geq 0
\ee
where the Shannon disorder of the copy is defined as in Eq.~(\ref{disorder}).  In this framework, the thermodynamic entropy production is again given by Eq.~(\ref{entr-prod-copolym}) but with the affinity\cite{AG08}
\be
A = - \frac{g}{k_{\rm B}T} +  D(\mbox{copy}\vert\mbox{template}) = - \frac{g}{k_{\rm B}T} + D(\mbox{copy}) - I(\mbox{copy},\mbox{template})
\label{Aff-template}
\ee
which establishes quantitatively a fundamental link between information and thermodynamics at the molecular scale.

\subsection{The case of DNA replication}

\begin{figure}
\centerline{\includegraphics[width=12cm]{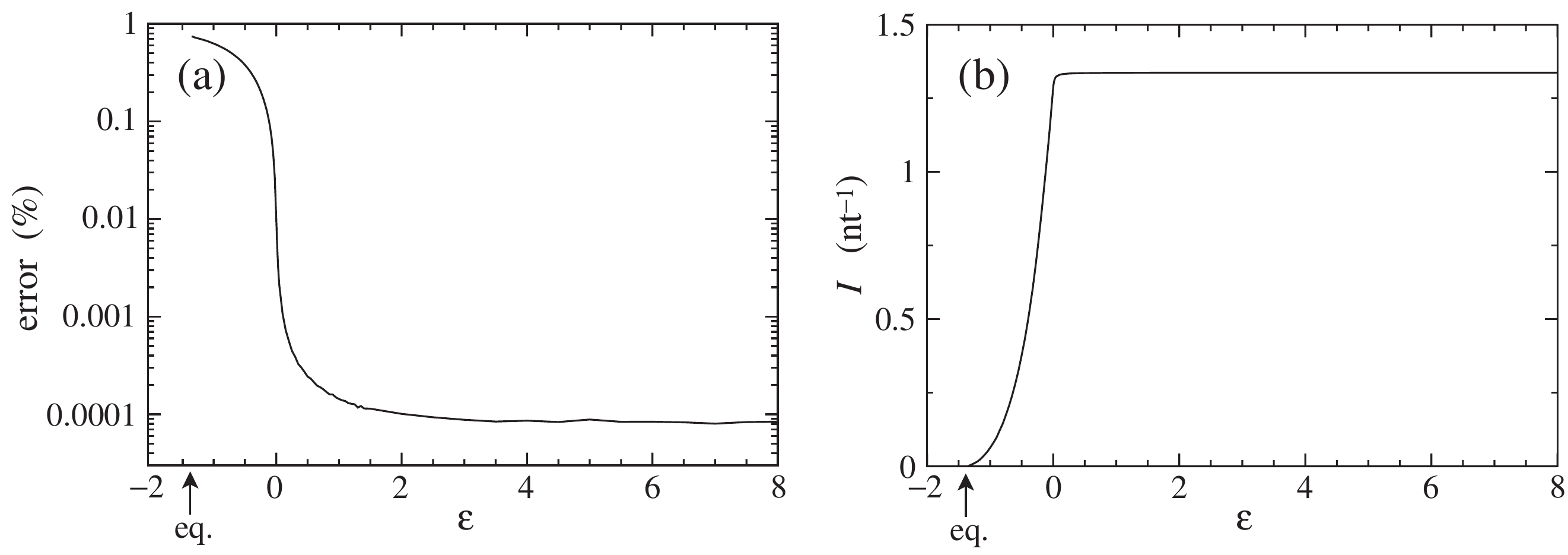}}
\caption{Stochastic simulation\cite{AG08} of DNA replication by polymerase Pol $\gamma$ of human mitochondrial DNA from GenBank\cite{GenBank} using known data on the kinetic constants of Watson-Crick pairing.\cite{LJ06} The reversed kinetic constants are taken as $k_{-mn}=k_{+mn}\,{\rm e}^{-\varepsilon}$. (a)~Percentage of errors in DNA replication versus the driving force $\varepsilon$. (b) Mutual information between the copied DNA strand and the original one versus the driving force.  The arrow points to the equilibrium value of the driving force: $\varepsilon_{\rm eq}=-\ln 4=-1.38629$.}
\label{fig6}
\end{figure}

The previous results apply to the different living copolymerization processes and, in particular, to DNA replication.  In this case, the subunits of the polymers are the four nucleotides N = A, T, C, or G and the monomers the corresponding nucleoside triphosphates NTP.  Assuming that no free energy difference exists between correct and incorrect chains, the copolymerization process can be simulated by Gillespie's algorithm as a function of the driving force $\varepsilon=-g/(k_{\rm B}T)$.\cite{AG08} The results are depicted in Fig.~\ref{fig6} which shows the percentage of replication errors as well as the mutual information between the copy and the template.  The error percentage is maximum at equilibrium and it decreases as the growth is pushed away from equilibrium.  Similarly, the mutual information vanishes at equilibrium and saturates at $I_{\rm max}=1.337$\;nats for high enough values of the driving force.  As in the case of free copolymerization, a transition occurs between the regime of growth by entropic effect for $\varepsilon_{\rm eq}<\varepsilon <0$ and the growth driven by free energy for $\varepsilon >0$.  At equilibrium, information transmission is not possible between the template and the copy.  Fidelity in the copying process becomes possible if enough free energy is supplied during copolymerization.

The existence of growth by entropic effect could be experimentally investigated in chemical or biological copolymerizations.  In polymer science, methods have not yet been much developed to perform the synthesis and sequencing of copolymers for the information they may support.  However, such methods are already well developed for DNA and under development for single-molecule DNA or RNA sequencing.\cite{GB06,Eid09,UEAKFTP10}  These methods could be used to test experimentally the predictions of copolymerization thermodynamics by varying NTP and pyrophosphate concentrations to approach the regime near equilibrium where the mutation rate increases.  

\section{Conclusions and perspectives}

Nowadays, self-organization has been studied for different phenomena from the macroscopic world down to the molecular scale.  At the macroscale, self-organization emerges far from equilibrium beyond bifurcations leading to bistability or oscillatory behavior.\cite{P67,NP77,IE95} However, on smaller and smaller scales, the time evolution of physico-chemical systems is more and more affected by thermal and molecular fluctuations which are the manifestation of the microscopic degrees of freedom.  In the framework of the theory of stochastic processes, the probability to find the system in some coarse-grained state is ruled by a master equation.  The macroscopic description in terms of deterministic kinetic equations is only obtained in the large-system limit, in which bifurcations emerge between nonequilibrium macrostates.  This emergence concerns in particular oscillations which only become correlated if the system is large enough.\cite{G02}  

This is the case for the nanoclocks of heterogeneous catalysis observed by field emission microscopy techniques such as the reaction of water formation on rhodium in high electric field under FIM conditions.\cite{MGVK09a,MGVK09b,MGVK10}   In the present contribution, methods are described for the modeling of these nonequilibrium processes at the nanoscale on highly non-uniform and anisotropic surfaces. The underlying crystalline structure determines the reactivity of the surface and contributes to the formation of the observed nanopatterns, which cannot be interpreted as standard reaction-diffusion patterns.  In spite of their nanometric size, such systems are large enough to undergo self-organization and manifest bistability and rhythmic behavior.

This dynamical order is the result of directionality which is induced away from equilibrium by the external constraints and free-energy sources characterized by the empowering affinities.  If the principle of detailed balancing holds at equilibrium, this is no longer the case out of equilibrium so that the fluctuating currents are biased and acquire a directionality, which is expressed by the current fluctuation theorem.\cite{AG04,AG06JSM,AG07JSP,AG07JSM,AG08JSM,AG09JSM,AGMT09}  This theorem has for consequence the non-negativity of the entropy production in accordance with the second law of thermodynamics, as well as generalizations of the Onsager reciprocity relations and Green-Kubo formula from the linear to the nonlinear response properties.\cite{AG04,AG07JSM}  Furthermore, the entropy production appears as a measure of time asymmetry in the temporal disorder of the typical histories of a system. Out of equilibrium, the typical histories are more probable than their time reversals, which appears as a corollary of the second law.\cite{G04JSP,G07CRP}  These new results transcend the known formulation of nonequilibrium thermodynamics, explicitly showing how the second law finds its origin in the breaking of time-reversal symmetry at the mesoscopic level of description in the theory of nonequilibrium systems.

These results show that the second law of thermodynamics already governs self-organization at the molecular scale.  Thermodynamics can be applied to the stochastic growth of a single copolymer, which is the support of information encoded in its sequence.  Away from equilibrium, dynamical order enables information processing during copolymerizations, which is not possible at equilibrium.  The statement by Manfred Eigen that ``information cannot originate in a system that is at equilibrium"\cite{E92} is rigorously proved in this framework.  The thermodynamics of copolymerization shows that, indeed, fundamental connections exist  between information and thermodynamics.\cite{AG08,J08,AG09}  In particular, the growth of a copolymer can be driven either by the free energy of the attachment of new monomers or by the entropic effect of disorder in the grown sequence.  

These considerations open new perspectives to understand the dynamical aspects of information in biology.  During copolymerization processes with a template as it is the case for replication, transcription or translation in biological systems, information is transmitted although errors may occur due to molecular fluctuations, which are sources of mutations.  Metabolism and self-reproduction, which are the two main features of biological systems, turn out to be linked in a fundamental way since information processing is constrained by energy dissipation during copolymerizations.  Moreover, the error threshold for the emergence of quasi-species in the hypercycle theory by Eigen and Schuster\cite{ES77-78} could be induced at the molecular scale by the transition towards high fidelity replication beyond the threshold at zero free energy per monomer between the two growth regimes.\cite{WW11} In this way, prebiotic chemistry could be more closely linked to the first steps of biological evolution.

\section*{Acknowledgements}
This research has been financially supported by the F.R.S.-FNRS, 
the ``Communaut\'e fran\c caise de Belgique'' (contract ``Actions de Recherche Concert\'ees'' No. 04/09-312), and the Belgian Federal Government (IAP project ``NOSY").


\begin{thebibliography}{9}

\bibitem{P67} I.~Prigogine, {\it Introduction to Thermodynamics of Irreversible Processes} (Wiley, New York, 1967).

\bibitem{NP77} G.~Nicolis and I.~Prigogine, {\it Self-Organization in Nonequilibrium Systems} (Wiley, New York, 1977). 

\bibitem{IE95} R.~Imbihl and G.~Ertl, {\it Chem. Rev.} {\bf 95}, 697 (1995).

\bibitem{NP89} G.~Nicolis and I.~Prigogine, {\it Exploring Complexity} (W.~H.~Freeman and Company, New York, 1989).

\bibitem{NN07} G.~Nicolis and C.~Nicolis, {\it Foundations of Complex Systems} (World Scientific, New Jersey, 2007).

\bibitem{FE11}  R. Feistel and W. Ebeling, {\it Physics of Self-Organization and Evolution} (Wiley-VCH, Weinheim, 2011).

\bibitem{G10} P. Gaspard, in: G. Radons, B. Rumpf, and H. G. Schuster, Editors,
{\it Nonlinear Dynamics of Nanosystems} (Wiley-VCH, Weinheim, 2010) pp. 1-74. 

\bibitem{vTGN92} M. F. H. van Tol, A. Gielbert, and B. E. Nieuwenhuys, {\it Catal. Lett.} {\bf 16}, 297 (1992).

\bibitem{GLRBE94} V. Gorodetskii, J. Lauterbach, H.-H. Rotermund, J. H. Block, and G. Ertl, {\it Nature} {\bf 370}, 276 (1994).

\bibitem{EBGKWB94} N. Ernst, G. Bozdech, V. Gorodetskii, H.-J. Kreuzer, R. L. C. Wang, and J.~H.~Block, {\it Surf. Sci.} {\bf 318}, L1211 (1994).

\bibitem{VK96} C. Voss and N. Kruse, {\it Appl. Surf. Sci.} {\bf 94}/{\bf 95}, 186 (1996).

\bibitem{KV08} N. Kruse and T. Visart de Bocarm\'e, in: G. Ertl, H. Kn\"ozinger, F. Sch\"uth, and J. Weitkamp, Editors, {\it Handbook of Heterogeneous Catalysis}, 2nd edition (Wiley-VCH, Weinheim, 2008) pp. 870-895.

\bibitem{VKGK06} T. Visart de Bocarm\'e, N. Kruse, P. Gaspard, and H. J. Kreuzer, {\it J. Chem. Phys.} {\bf 125}, 054704 (2006).

\bibitem{MG06} J.-S. McEwen and P. Gaspard, {\it J. Chem. Phys.} {\bf 125}, 214707 (2006).

\bibitem{MGMVK08} J.-S. McEwen, P. Gaspard, F. Mittendorfer, T. Visart de Bocarm\'e, and N.~Kruse, {\it Chem. Phys. Lett.} {\bf 452}, 133 (2008).

\bibitem{VBK01} T. Visart de Bocarm\'e, T. B\"ar, and N. Kruse, {\it Ultramicroscopy} {\bf 89}, 75 (2001).

\bibitem{VBK04} T. Visart de Bocarm\'e, G. Beketov, and N. Kruse, {\it Surf. Interface Anal.} {\bf 36}, 522 (2004).

\bibitem{MGVK09a} J.-S. McEwen, P. Gaspard, T. Visart de Bocarm\'e, and N. Kruse, {\it Proc. Natl. Acad. Sci. USA} {\bf 106}, 3006 (2009).

\bibitem{MGVK09b} J.-S. McEwen, P. Gaspard, T. Visart de Bocarm\'e, and N. Kruse, {\it J. Phys. Chem. C} {\bf 113}, 17045 (2009).

\bibitem{MGVK10} J.-S. McEwen, P. Gaspard, T. Visart de Bocarm\'e, and N. Kruse, {\it Surf. Sci.} {\bf 604}, 1353 (2010).

\bibitem{AG08} D. Andrieux and P. Gaspard, {\it Proc. Natl. Acad. Sci. USA} {\bf 105}, 9516 (2008).

\bibitem{J08} C. Jarzynski, {\it Proc. Natl. Acad. Sci. USA} {\bf 105}, 9451 (2008).

\bibitem{AG09} D. Andrieux and P. Gaspard, {\it J. Chem. Phys.} {\bf 130}, 014901 (2009).

\bibitem{GMK10} V. Garcia-Morales and K. Krischer, {\it Proc. Natl. Acad. Sci. USA} {\bf 107}, 4528 (2010).

\bibitem{SG11} J.-P. Sauvage and P. Gaspard, Editors, {\it From Non-Covalent Assemblies to Molecular Machines} (Wiley-VCH, Weinheim, 2011).

\bibitem{LSBFLD11} P. Lussis, T. Svaldo-Lanero, A. Bertocco, C.-A. Fustin, D. A. Leigh, and A.-S. Duwez, {\it Nature Nanotech.} {\bf 6}, 553 (2011).

\bibitem{MJYKQX05} W. Min, L. Jiang, J. Yu, S. C. Kou, H. Qian, and X. S. Xie, {\it Nano Lett.} {\bf 5}, 2373 (2005).

\bibitem{FHTSY95} T. Funatsu, Y. Harada, M. Tokunaga, K. Saito, and T. Yanagida, {\it Nature} {\bf 374}, 555 (1995).

\bibitem{NYYK97} H. Noji, R. Yasuda, M. Yoshida, and K. Kinosita Jr., {\it Nature} {\bf 386}, 299 (1997).

\bibitem{ALSLRW02} K. Adelman, A. La Porta, T. J. Santangelo, J. T. Lis, J. W. Roberts, and M. D. Wang, {\it Proc. Natl. Acad. Sci. USA} {\bf 99}, 13538 (2002).

\bibitem{GB06} W. J. Greenleaf and S. M. Block, {\it Science} {\bf 313}, 801 (2006).

\bibitem{Eid09} J. Eid {\it et al.}, {\it Science} {\bf 323}, 133 (2009).

\bibitem{UEAKFTP10} S. Uemura, C. Echeverr\'{\i}a Aitken, J. Korlach, B. A. Flusberg, S. W. Turner, and J. D. Puglisi, {\it Nature} {\bf 464}, 1012 (2010).

\bibitem{AG06PRE} D. Andrieux and P. Gaspard, {\it Phys. Rev. E} {\bf 74}, 011906 (2006).

\bibitem{GG10} E. Gerritsma and P. Gaspard, {\it Biophys. Rev. Lett.} {\bf 5}, 163 (2010).

\bibitem{GN90} P. Gaspard and G. Nicolis, {\it Phys. Rev. Lett.} {\bf 65}, 1693 (1990).

\bibitem{G05} P. Gaspard, {\it New J. Phys.} {\bf 7}, 77 (2005).

\bibitem{ECM93} D.~J.~Evans, E.~G.~D.~Cohen, and G.~P.~Morriss, {\it Phys. Rev. Lett.} {\bf 71}, 2401 (1993).

\bibitem{GC95} G.~Gallavotti and E.~G.~D.~Cohen, {\it Phys. Rev. Lett.} {\bf 74}, 2694 (1995).

\bibitem{C99} G. E. Crooks, {\it Phys. Rev. E} {\bf 60}, 2721 (1999).

\bibitem{LS99} J.~L.~Lebowitz and H.~Spohn, {\it J. Stat. Phys.} {\bf 95}, 333 (1999).

\bibitem{EHM09} M.~Esposito, U.~Harbola, and S.~Mukamel, {\it Rev. Mod. Phys.} {\bf 81}, 1665 (2009).

\bibitem{J11} C. Jarzynski, {\it Annu. Rev. Condens. Matter Phys.} {\bf 2}, 329 (2011).

\bibitem{CRJSTB05} D. Collin, F. Ritort, C. Jarzynski, S. B. Smith, I. Tinoco Jr., and C. Bustamante,
{\it Nature} {\bf 437}, 231 (2005).

\bibitem{AG04} D. Andrieux and P. Gaspard, {\it J. Chem. Phys.} {\bf 121}, 6167 (2004).

\bibitem{AG06JSM} D. Andrieux and P. Gaspard, {\it J. Stat. Mech.: Th. Exp.} P01011 (2006).

\bibitem{AG07JSP} D. Andrieux and P. Gaspard, {\it J. Stat. Phys.} {\bf 127}, 107 (2007).

\bibitem{AG07JSM} D. Andrieux and P. Gaspard, {\it J. Stat. Mech.: Th. Exp.} P02006 (2007).

\bibitem{AG08JSM} D. Andrieux and P. Gaspard, {\it J. Stat. Mech.: Th. Exp.} P11007 (2008).

\bibitem{AG09JSM} D. Andrieux and P. Gaspard, {\it J. Stat. Mech.: Th. Exp.} P02057 (2009).

\bibitem{AGMT09} D. Andrieux, P. Gaspard, T. Monnai, and S. Tasaki, {\it New J. Phys.} {\bf 11}, 043014 (2009).

\bibitem{DD36} T. De Donder and P. Van Rysselberghe, {\it Affinity} (Stanford University Press, Menlo Park CA, 1936).

\bibitem{AG08JCP} D. Andrieux and P. Gaspard, {\it J. Chem. Phys.} {\bf 128}, 154506 (2008).

\bibitem{GW93} P. Gaspard and X.-J. Wang, {\it Phys. Rep.} {\bf 235}, 291 (1993).

\bibitem{G98} P. Gaspard, {\it Chaos, Scattering and Statistical Mechanics}
(Cambridge University Press, Cambridge UK, 1998).

\bibitem{G04JSP} P. Gaspard, {\it J. Stat. Phys.} {\bf 117}, 599 (2004).

\bibitem{CT06} T. M. Cover and J. A. Thomas, {\it Elements of Information Theory}, 2nd edition (Wiley, Hoboken, 2006).

\bibitem{AGCGJP07} D. Andrieux, P. Gaspard, S. Ciliberto, N. Garnier,
S. Joubaud, and A. Petrosyan, {\it Phys. Rev. Lett.} {\bf 98}, 150601 (2007).

\bibitem{AGCGJP08} D. Andrieux, P. Gaspard, S. Ciliberto, N. Garnier,
S. Joubaud, and A. Petrosyan, {\it J. Stat. Mech.: Th. Exp.} P01002 (2008).

\bibitem{G07CRP} P. Gaspard, {\it C. R. Physique} {\bf 8}, 598 (2007).

\bibitem{MT69} E. W. M\"uller and T. T. Tsong, {\it Field Ion Microscopy: Principles and Applications} (Elsevier, New York, 1969).

\bibitem{ME95} A. S. Mikhailov and G. Ertl, {\it Chem. Phys. Lett.} {\bf 238}, 104 (1995).

\bibitem{HME98} M. Hildebrand, A. S. Mikhailov, and G. Ertl, {\it Phys. Rev. E} {\bf 58}, 5483 (1998).

\bibitem{DMHGKMI04} Y. De Decker, A. Marbach, M. Hinz, S. G\"unther, M. Kiskinova, A.~S.~Mikhailov, and R. Imbihl, {\it Phys. Rev. Lett.} {\bf 92}, 198305 (2004).

\bibitem{GMG11} A. Garc\'{\i}a Cant\'u Ros, J.-S. McEwen, and P. Gaspard, {\it Phys. Rev. E} {\bf 83}, 021604 (2011).

\bibitem{SLWC90} E. Schwarz, J. Lenz, H. Wohlgemuth, and K. Christmann, {\it Vacuum} {\bf 41}, 167 (1990).

\bibitem{SS93} A. N. Salanov and V. I. Savchenko, {\it Surf. Sci.} {\bf 296}, 393 (1993).

\bibitem{GRS02} M. V. Ganduglia-Pirovano, K. Reuter, and M. Scheffler, {\it Phys. Rev. B} {\bf 65}, 245426 (2002).

\bibitem{GMBLKKSVYTQA04} J. Gustafson, A. Mikkelsen, M. Borg, E. Lundgren, L. K\"ohler, G. Kresse, M. Schmid, P. Varga, J. Yuhara, X. Torrelles, C. Quir\'os, and J. N. Andersen, {\it Phys. Rev. Lett.} {\bf 92}, 126102 (2004).

\bibitem{AELCHRC04} C. Africh, F. Esch, W. X. Li, M. Corso, B. Hammer, R. Rosei, and G. Comelli, {\it Phys. Rev. Lett.} {\bf 93}, 126104 (2004).

\bibitem{KKSLGMBYAMV04} L. K\"ohler, G. Kresse, M. Schmid, E. Lundgren, J. Gustafson, A. Mikkelsen, M. Borg, J. Yuhara, J. N. Andersen, M. Marsman, and P. Varga, {\it Phys. Rev. Lett.} {\bf 93}, 266103 (2004).

\bibitem{GMBALKHSVKKKSD05} J. Gustafson, A. Mikkelsen, M. Borg, J. N. Andersen, E. Lundgren, C. Klein, W. Hofer, M. Schmid, P. Varga, L. K\"ohler, G. Kresse, N. Kasper, A. Stierle, and H. Dosch, {\it Phys. Rev. B} {\bf 71}, 115442 (2005).

\bibitem{DAECDKMKDK06} C. Dri, C. Africh, F. Esch, G. Comelli, O. Dubay, L. K\"ohler, F. Mittendorfer, G. Kresse, P. Dubin, and M. Kiskinova, {\it J. Chem. Phys.} {\bf 125}, 094701 (2006).

\bibitem{LMAKSV06} E.~Lundgren, A.~Mikkelsen, J.~N.~Andersen, G.~Kresse, M.~Schmid, and P.~Varga, {\it J. Phys.: Condens. Matter} {\bf 18}, R481 (2006).

\bibitem{M10} F. Mittendorfer, {\it J. Phys.: Condens. Matter} {\bf 22}, 393001 (2010).

\bibitem{MGDBVK10} J.-S. McEwen, P. Gaspard, Y. De Decker, C. Barroo, T. Visart de Bocarm\'e, and N. Kruse, {\it Langmuir} {\bf 26}, 16381 (2010).

\bibitem{G02} P. Gaspard, {\it J. Chem. Phys.} {\bf 117}, 8905 (2002).

\bibitem{S44} E. Schr\"odinger, {\it What is Life?} (Cambridge University Press, Cambridge UK, 1944).

\bibitem{CF63JPS} B. D. Coleman and T. G. Fox, {\it J. Polym. Sci. A} {\bf 1}, 3183 (1963).

\bibitem{CF63JCP} B. D. Coleman and T. G. Fox, {\it J. Chem. Phys.} {\bf 38}, 1065 (1963).

\bibitem{GenBank} Homo sapiens DNA mitochondrion, 16569 bp, locus AC 
000021, version GI:115315570, {\tt http://www.ncbi.nlm.nih.gov}.

\bibitem{LJ06} H. Lee and K. Johnson, {\it J. Biol. Chem.} {\bf 281}, 36236 (2006).

\bibitem{E92} M. Eigen, {\it Steps towards Life: A Perspective on Evolution} (Oxford University Press, Oxford, 1992).

\bibitem{ES77-78} M. Eigen and P. Schuster, {\it Naturwissenschaften} {\bf 64}, 541 (1977); {\it ibid.} {\bf 65}, 7 (1978); {\it ibid.} {\bf 65}, 341 (1978).

\bibitem{WW11} H.-J. Woo and A. Wallqvist, {\it Phys. Rev. Lett.} {\bf 106}, 060601 (2011).

\end{thebibliography}
\end{document}